\documentclass[DIV15,12pt,a4paper,british]{scrartcl}
\usepackage[T1]{fontenc}
\usepackage[active]{srcltx}
\usepackage{calc}
\usepackage{units}
\usepackage{url}
\usepackage{amstext}
\usepackage{amssymb}
\usepackage{graphicx}
\usepackage{cuted}
\usepackage[normalem]{ulem}
\usepackage{cancel}
\usepackage{subfig}
\makeatletter

\usepackage{layout}

\makeatother

\begin{document}

\title{Requirements and Design Methodology for a Spatio-Angular Vitreoretinal Surgical Microscope}

\author{Brice Thurin and Christos Bergeles}

\publishers{Wellcome/EPSRC Centre for Interventional and Surgical Sciences, Department of Medical Physics and Biomedical Engineering, UCL, London, UK \\ 
Centre for Medical Image Computing, Department of Medical Physics and Biomedical Engineering, UCL, London, UK\\
Corresponding author: brice@aeptika.co.uk}

\date{Compiled \today}

\maketitle
\begin{abstract}
This paper presents the requirements and design methodology for a vitreoretinal surgical microscope based on plenoptic imaging. The design parameters of the imaging lens, micro-lens array, and sensor are specified. The proposed design provides an extended depth imaging range when compared to proof-of-concept systems based on a commercial plenoptic sensor, and serves as the guideline for the implementation of clinically relevant surgical microscopes.
\end{abstract}
\section{Introduction}
\label{sec:intro}

Vitreoretinal surgery (VRS) takes place under high-magnification stereo microscopy that limits the attainable depth of the imaging volume, hinders $3$D perception, and requires constant manual adjustment of microscope focus \cite{mohamed_first_2017}. Towards improving imaging in VRS, we propose retinal observation via a plenoptic sensor, \emph{i.e.}~a plenoptic surgical microscope. An implementation of a plenoptic, or ``light-field'', sensor is a photodetector array with a micro-lens array in front of it \cite{zappe_plenoptic_2016,ihrke_principles_2016}. The micro-lenses can be considered as an array of micro-cameras, each of which captures slightly overlapping micro-images. The parallax between the micro-images allows $3$D reconstruction and computational refocusing of images and video streams \cite{lam_computational_2015}.

Several papers and patents applications \cite{adam_plenoptic_2016, tumlinson_light_2011,bedard_simultaneous_2014,lawson_methods_2014,abt_ophthalmic_2015} showcase the interest in this imaging modality, but lack a design-driven implementation or considerations to improve image quality. The technology was evaluated as a diagnostic tools to assess the health of the human iris \emph{in vivo}\cite{chen_human_2017}.

\begin{figure}[tbph]
\begin{center}
	\fbox{\includegraphics[width=\linewidth]{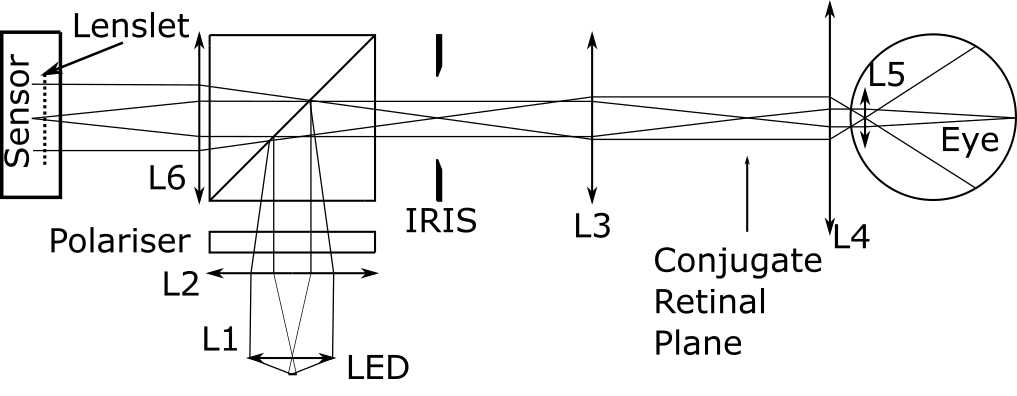}}
\end{center}
\caption{Layout of the optics of the engineered fundus camera.}
\label{fig:optical_layout}
\end{figure}

\begin{figure}[tbph]
\begin{center}
	\fbox{\includegraphics[width=\linewidth]{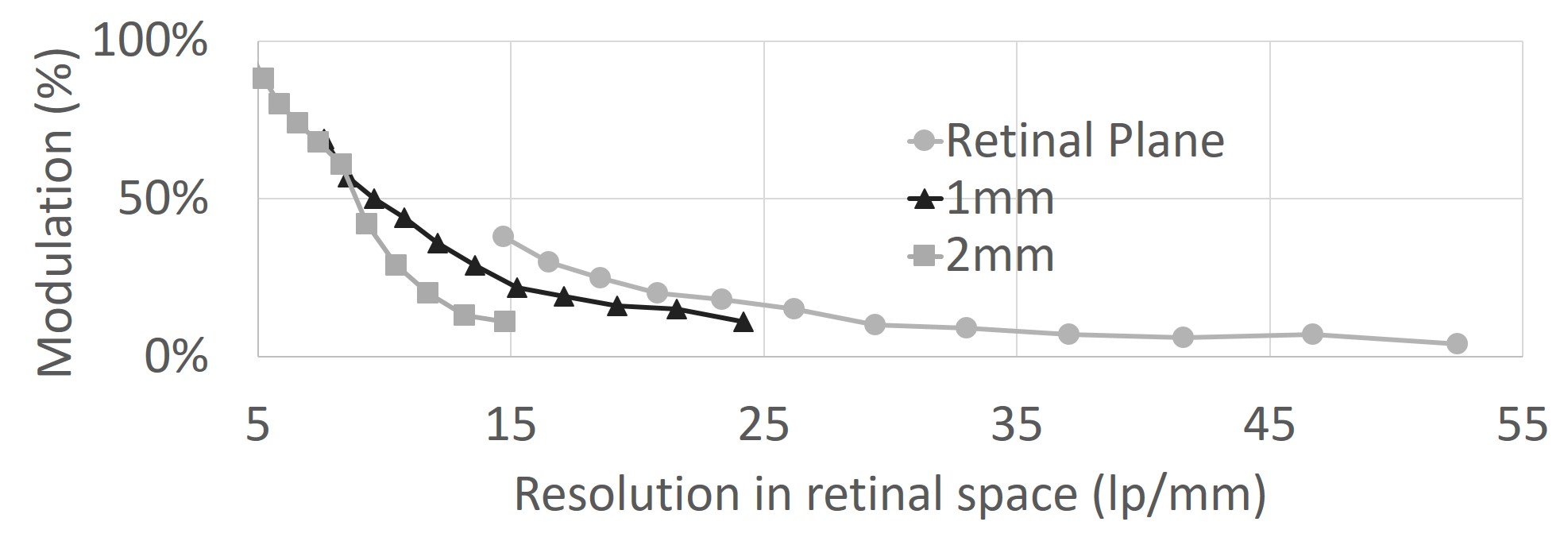}}
\end{center}
\caption{Measured Modulation Transfer Function for the Plenoptic Camera at different depth in retinal space.}
\label{fig:measured_mtf}
\end{figure}

\begin{figure}[tbph]
\centering
\fbox{\includegraphics[width=\linewidth]{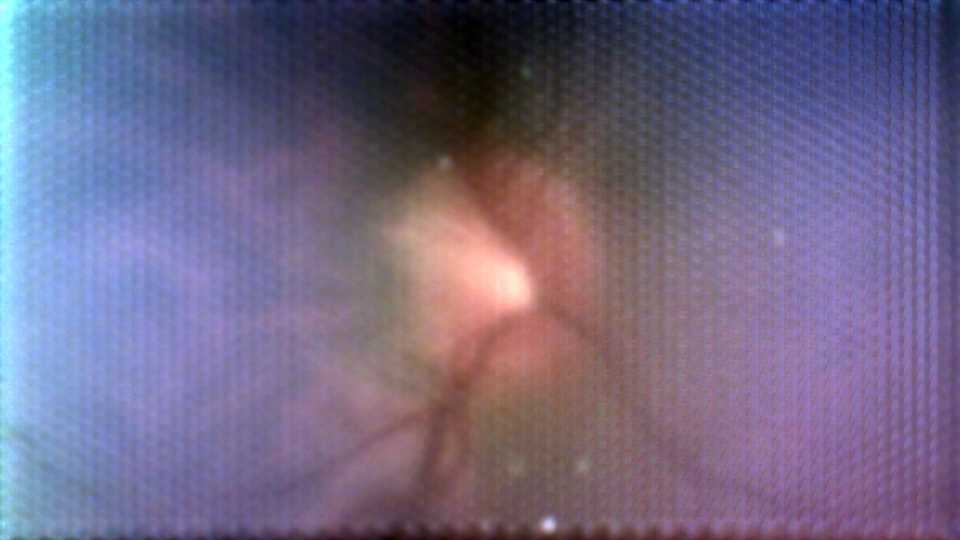}}
\caption{\emph{In vivo} digitally refocused image of the optic disc of a myopic volunteer without dioptric correction.}
\label{fig:invivo}
\end{figure}

Recently, we implemented a proof-of-concept stand-alone plenoptic ophthalmoscope \cite{thurin_retinal_2018}. Its optical layout is given in Fig.\ \ref{fig:optical_layout}. We used the R$8$ multi-focus plenoptic sensor from Raytrix GmbH \cite{perwass_single_2012}, which has $3$ sets of interlaced micro-lenses, each with a different focal length. The prototype' performance was measured following the ISO standard method\cite{noauthor_ophthalmic_nodate-1} of imaging printed $1951$ USAF Resolution Targets. Figure\ \ref{fig:measured_mtf} depicts the Modulation Transfer Function~(MTF) normalised at $0\,\nicefrac{\textnormal{lp}}{\textnormal{mm}}$ as a function of resolution of the different set bars present on the chart for the target located at a retinal plane conjugate~($\circ$), $1\,$mm~($\triangle$) and $2\,$mm~($\square$) away from the retina. For the plane $2\,$mm~away from the retina the plenoptic camera can just about resolve $15\,\nicefrac{\textnormal{lp}}{\textnormal{mm}}$ features. We observed that the lateral resolution decreases rapidly for imaging planes further from the retina. We concluded our proof-of-concept evaluation by imaging a phantom eye with a surgical tool, and a human eye \emph{in vivo}. Figure\ \ref{fig:invivo} shows an image of the optic disc of a $-3\,$Diopters myopic volunteer, acquired without manual focusing adjustment and computationally refocused post-capture.

In this letter, we present a design framework that address the limitations of this proof-of-concept system. Notably we specify the design parameters that maximize the imaging depth range of an ophthalmic plenoptic retinal imaging system without decrease in lateral resolution.

\section{Design Considerations}
\label{sec:design-considerations}

The primary design principle for plenoptic imaging is to match the
effective f-number of a micro-lens, $f_{\#}$, and of the main
lens, $F_{\#}$,~\cite{turola_investigation_2016}. This constraint maximises the fill factor of the sensor while avoiding aliasing caused by overlapping micro-images. 
    
As imaging resolution improves with smaller pupil diameters due to reduced ocular optical aberrations \cite{watson_computing_2015}, standard retinal imaging systems limit the optical aperture to a portion of the pupil less than $3\,$mm in diameter~\cite{dehoog_novel_2009}. To achieve high axial resolution via plenoptic imaging, however, it is necessary to maximize the parallax among the micro-images. Therefore, the entrance aperture of the plenoptic imaging system is better set by the eye pupil's physical size. 

Perwass \emph{et al.} \cite{perwass_single_2012} introduced the concept of ``virtual depth'' as the ratio of the distance between the main lens image plane and the micro-lens array, $a$, and the distance between the sensor plane and the micro-lens array plane, $b$, (see Fig.\ \ref{fig:lfo_model}). While $b$ is fixed by the plenoptic sensor assembly, $a$ covers the range of image plane depths for which the blur spot on the sensor is smaller than the pixel size. If $d$ is the micro-lens' diameter: 
\begin{equation}
	v = \frac{a}{b} = \frac{a}{d f_{\text{\#}}},\label{eq:virtual depth}
\end{equation}
which, given a range of virtual depth values $v$ resolvable by the plenoptic sensor constrains focusing within $\left[a_{min},\,a_{max}\right]$.

In general, the effective imaging lens f-number of the main lens is dependent on the image distance. Therefore the matching f-number condition introduced earlier will not hold for the whole depth imaging range of the plenoptic sensor. However if the imaging lens is telecentric in imaging space, i.e. the eye pupil plane is conjugated with the front focal plane of the imaging lens, the effective imaging lens f-number ($F_{\#}$) remains constant and is given by the ratio of the effective focal length of the main lens, $F$, to the entrance pupil diameter, \emph{i.e.} the diameter of the entire eye pupil, $\Delta$, hence e $f_{\#}=F_{\#}=\nicefrac{F}{\Delta}$.
\section{Plenoptic Imaging System Model}

\begin{figure}[tbh]
\centering
\subfloat[Retinal plane conjugate]{\fbox{\includegraphics[width=0.99\linewidth]{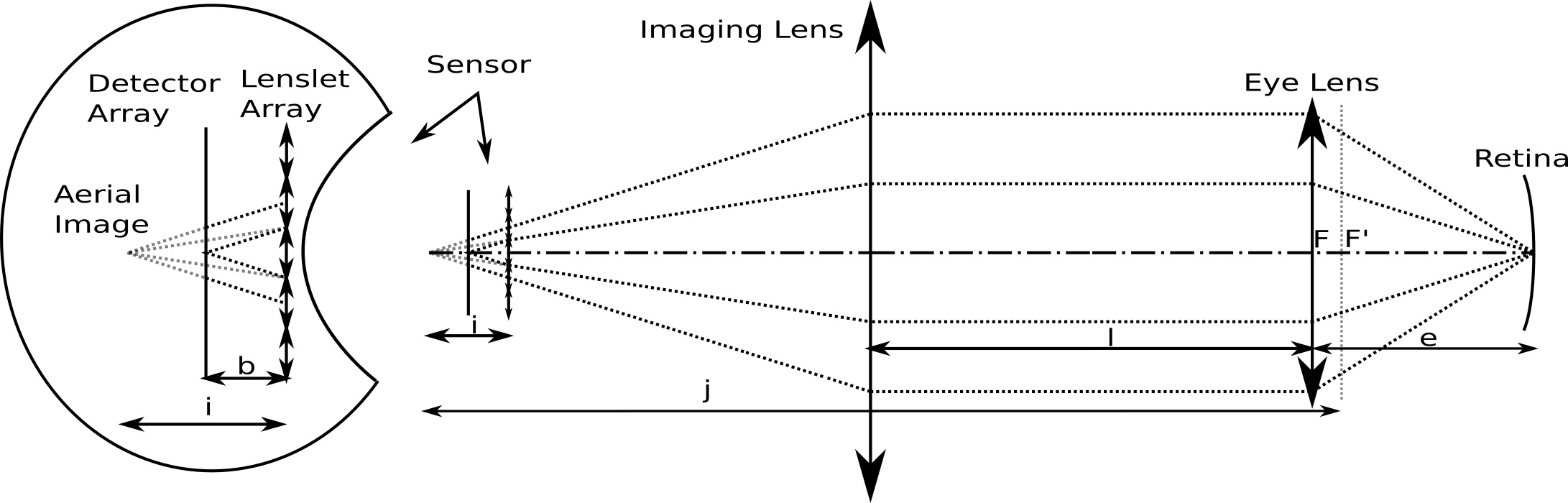}}}
\vfill
\subfloat[Alternative depth plane]{\fbox{\includegraphics[width=0.99\linewidth]{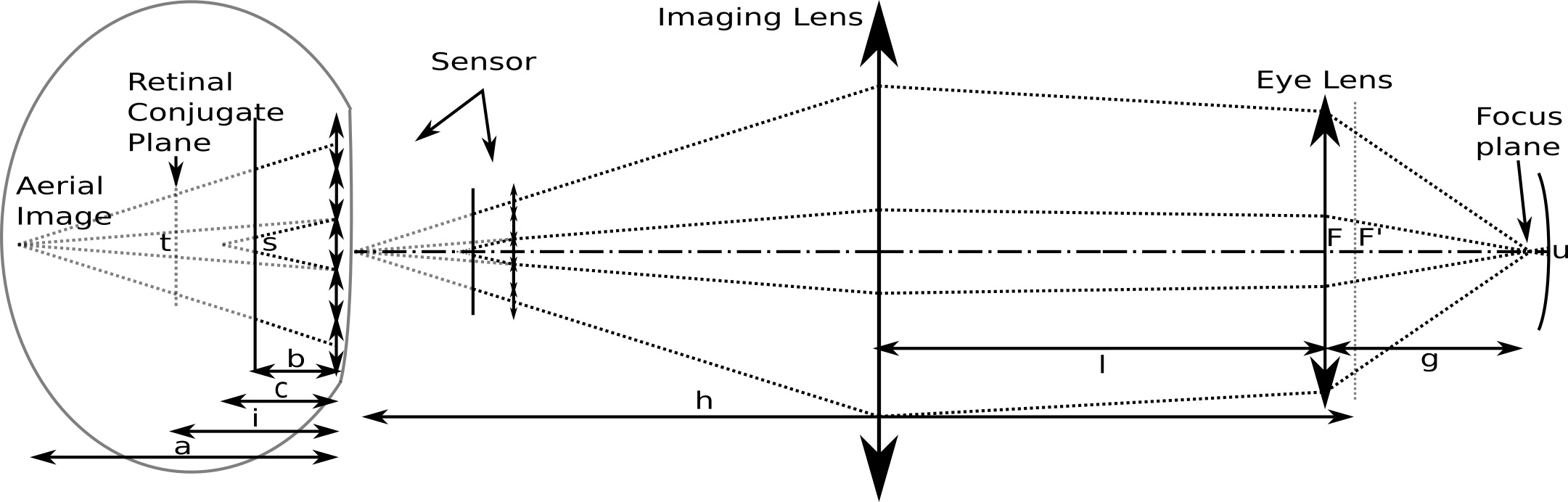}}}
\caption{Model for design parameters estimation.}
\label{fig:lfo_model}
\end{figure}

A single lens is used to model the optics of the eye in front of the retina. It is well known that the eye's optics can be well approximated by a single lens, which merely ignores some of the higher-order aberrations. An imaging lens creates an aerial image of the retina. The micro-lens array projects multiple non-overlapping images of the retina on top of the photodetector array. Figure\ \ref{fig:lfo_model}(a) shows the layout of the model with rays traced from the retinal plane coming into focus on the detector array. Figure\ \ref{fig:lfo_model}(b) shows the same layout but for rays emanating from a plane in front of the retina, for which the images on the detector are slightly out of focus. The imaging lens is located a distance $l=f^L$ away from the eye to satisfy the telecentric requirements previously introduced.

The virtual image of the retina formed by the combined imaging and eye lenses is brought in focus by the micro-lenses onto the sensor. For planes at different depths, the dimension of the blur spot on the sensor, $s_{0p}$, and its projection, $u_{0p}$, in the retinal space are used to establish the depth-of-field (for clarity we omit the subscript in the figure).

As illustrated in Fig.\ \ref{fig:lfo_model}, $s_{0p}$ (indicated as $s$) corresponds to a plane being focused at a distance $c_{p}$ from the micro-lens array, where $p\in\left[0,n-1\right]$ corresponds to each of $n$ different types (different focal lengths) of interlaced micro-lenses. Using similar triangles, the blur spot $s_{0p}$ can be expressed as a function of the micro-lens diameter, $d$, and the distance between the sensor and micro-lens array, $b$:
\begin{equation}
\frac{s_{0p}}{d}=\frac{c_{p}-b}{c_{p}}.\label{eq:s}
\end{equation}
The thin-lens equations for a micro-lens are
\begin{equation}
\frac{1}{f_{p}} = \frac{1}{c_{p}}-\frac{1}{a}\,\textnormal{, and}\,\frac{1}{f_{p}}=\frac{1}{b}-\frac{1}{i_{p}}.\label{eq:lenslet thin lens}
\end{equation}
The f-number-matching condition is written as
\begin{equation}
\frac{b}{d}=\frac{F}{\Delta}.\label{eq:fnumber matching}
\end{equation}
In the telecentric configuration proposed in this letter, the effective focal length of the eye and imaging lens combination is equal to the imaging lens focal length $f^L$. In Fig.\ \ref{fig:lfo_model}, F and F' represents the object and image foci respectively, the Newtonian lens equation for the Imaging lens and eye lens combination are:
\begin{equation}
gh=-F^{2}\ \textnormal{ and }\ e_{p}j_{p}=-F^{2}
\end{equation}
with the equality defining the spacing between the image focus F' and the micro-lens array:
\begin{equation}
h-a=j_{0}-i_{0}.\label{eq:lens lenslet image distance}
\end{equation}
We also introduce $s_{\lambda}$ as the sensor resolution limit which is equal to twice the pixel size. The blur spot $u_{0p}$ is given by the inverse magnification of the micro-lens array and the eye and imaging lens combination:
\hspace{-4pt}

\begin{eqnarray}
u_{0p}&=&\Bigg\vert\frac{s_{0p}e_{p}i_{p}}{Fb}\Bigg\vert = \left(1-\frac{1}{\left(\frac{1}{\frac{1}{b}-\frac{1}{f_{0}}}+\left(\frac{1}{e_{0}}-\frac{1}{g}\right)\left(\frac{\Delta b}{d}\right)^{2}\right)\left(\frac{1}{b}-\frac{1}{f_{p}}\right)}\right)\frac{\Delta b}{\left(\frac{-\Delta^{2}b^{2}}{d^{2}e_{0}}-\frac{1}{\frac{1}{b}-\frac{1}{f_{0}}}+\frac{1}{\frac{1}{b}-\frac{1}{f_{p}}}\right)} \label{eq:u} \\
u_{\lambda} & = & \Bigg\vert\frac{s_{\lambda}ag}{bF}\Bigg\vert =\left(\left(k-e_{0}\right)\left(\frac{1}{\frac{1}{b}-\frac{1}{f_{0}}}\frac{d}{\Delta b^{2}}+\frac{1}{e_{0}}\frac{\Delta}{d}\right)+\frac{\Delta}{d}\right)\label{eq:ulambda}
\end{eqnarray}

\hspace{-4pt}where we introduced variables $k=e_{0}-g$, with $k,\,e_{0},\,g<0$. Hence, we have $2$ variables: $e_{0}$ defining the reference retinal plane ($e_{0}=-17$\,mm), and $k$ defining the distance from $e_{0}$. 

The plenoptic sensor design parameters $b,$ $d$, and $f_{p}$, must be optimized so that the blur spot diameter, $u=\mathbf{\max}\left[\mathbf{\min}\left[u_{00},\ldots,u_{0n}\right],u_{\lambda}\right]$, is smaller than the resolution target value. The ISO standards require a resolution of $60\,\nicefrac{\textnormal{lp}}{\textnormal{mm}}$ for camera with a field of view $>30^o$ , which results in $u<8.33\,\mu$m. The design problem pertains to finding the parameters that provide the largest (absolute) value of $k$ such that $u$ fulfills the ISO directives.

The multi-focus plenoptic sensor is composed of regularly spaced interlaced micro-lenses arrays of different focal lengths. When only a single micro-lens type is used, \emph{i.e.}~$n=1$, the minimum - \emph{double covering} - virtual distance required to image any point along any direction by at least two micro-lenses is equal
to $v_{min}=2.3$~\cite{perwass_single_2012} for an hexagonal array. For $n=3$, the \emph{double covering}
virtual distance is $v_{min}=4$~\cite{perwass_single_2012}. Finally for the next two regularly interlaced hexagonal arrays, $n=4$, $v_{min}=4.6$  and $n=7$, $v_{min}\approx6.1$. The maximum achievable resolution in the reference plane is calculated from (\ref{eq:ulambda}) for $k=0$:
\begin{equation}
u_{\lambda}\left(k=0\right)=s_{\lambda}\frac{\Delta}{d}.\label{eq:ulambda vmin}
\end{equation}

\begin{figure}[tbph]
\centering
\fbox{\includegraphics[width=\linewidth]{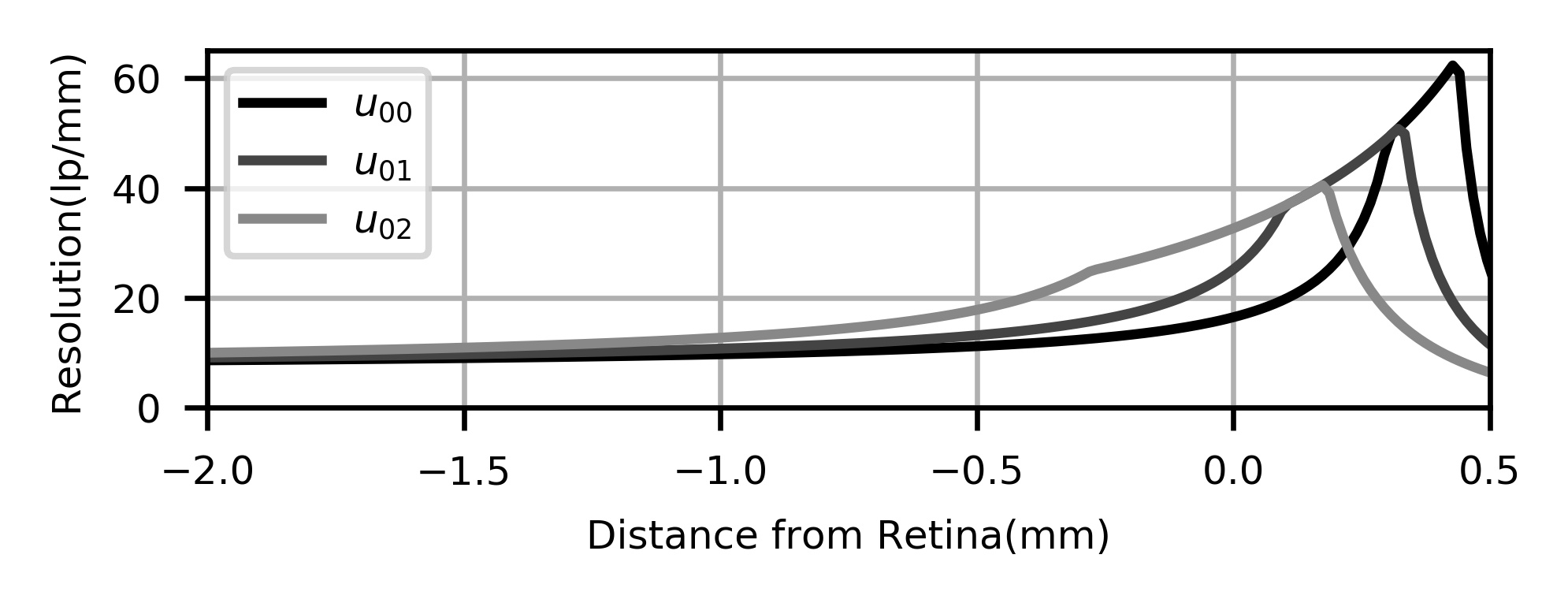}}
\caption{Resolution for each of the $3$ micro-lens types of Raytrix R$8$ sensor.}
\label{fig:R8_u_res}
\end{figure}

The resolution estimated from our model for the R$8$ sensor
is shown in Fig.\ \ref{fig:R8_u_res}. The theoretical resolution
limit quickly reduces for object planes located away from the retina within the vitreous. For plane just $0.5$\,mm away from the retina the resolution is less than $20\,\nicefrac{\textnormal{lp}}{\textnormal{mm}}$, which is in agreement with the MTF measurements reported in Sec.\ \ref{sec:intro}.

\begin{figure}[tbph]
\centering
\fbox{\includegraphics[width=\linewidth]{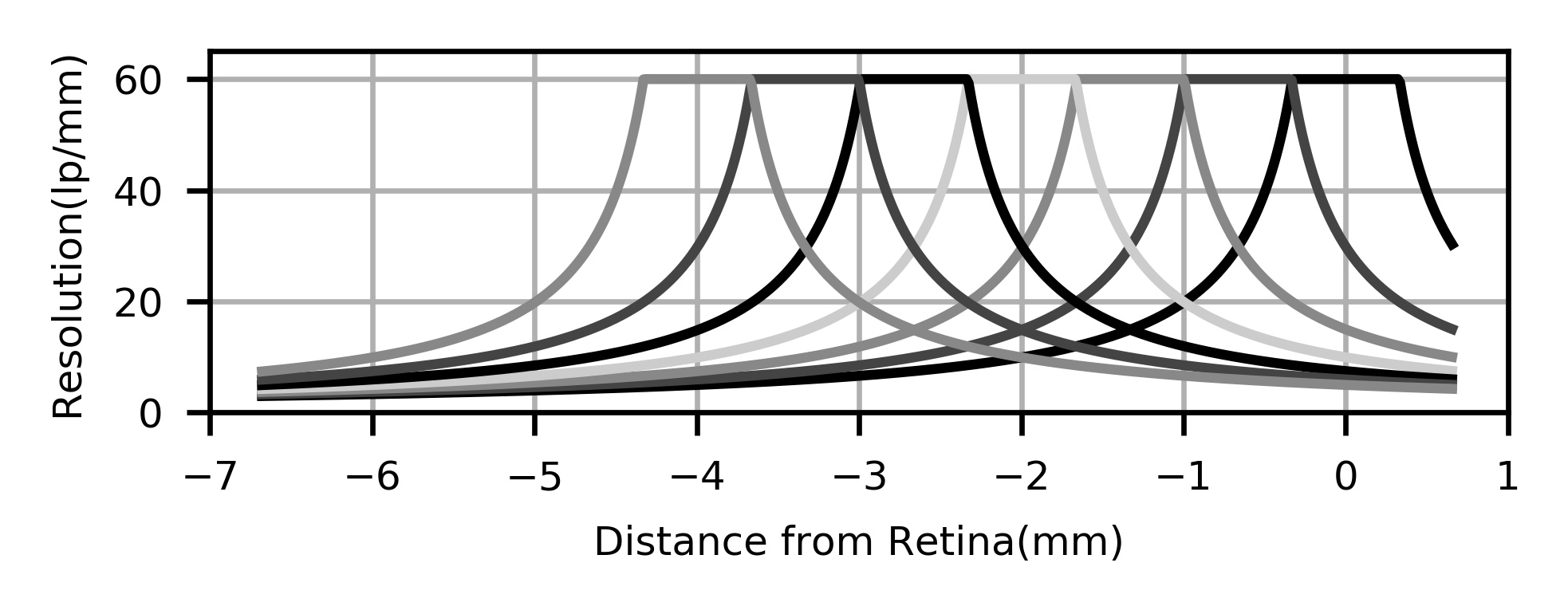}}
\caption{Resolution for $7$ interlaced micro-lens types on a hexagonal array with a lateral resolution constant within the imaging volume.}
\label{fig:7 lenslet f3 nodiff}
\end{figure}

Considering (\ref{eq:ulambda}), the diffraction-limit spot size can be made independent of $k$, if the distance, $b$, from the micro-lens array to the sensor is, with $v_{min}=\nicefrac{i}{b}$:
\begin{equation}
b=-e_{0}v_{min}\frac{d^{2}}{\Delta^{2}}.\label{eq:minimizer}
\end{equation}
In that case, it can be shown that (\ref{eq:ulambda}) reduces to (\ref{eq:ulambda vmin}). Using the condidtion of equation\ \ref{eq:minimizer} into our model, we display in figure\ \ref{fig:7 lenslet f3 nodiff}, the resolution curve we expect to get with a 7-types hexagonal lenslet array. We select the design parameter of the plenoptic sensor in order to achieve a constant lateral resolution of $60\,\nicefrac{\textnormal{lp}}{\textnormal{mm}}$ for an imaging volume with a depth of more than $4\,\textnormal{mm}$.
\section{Discussion}
\begin{figure}[tbph]
\centering
\fbox{\includegraphics[width=\linewidth]{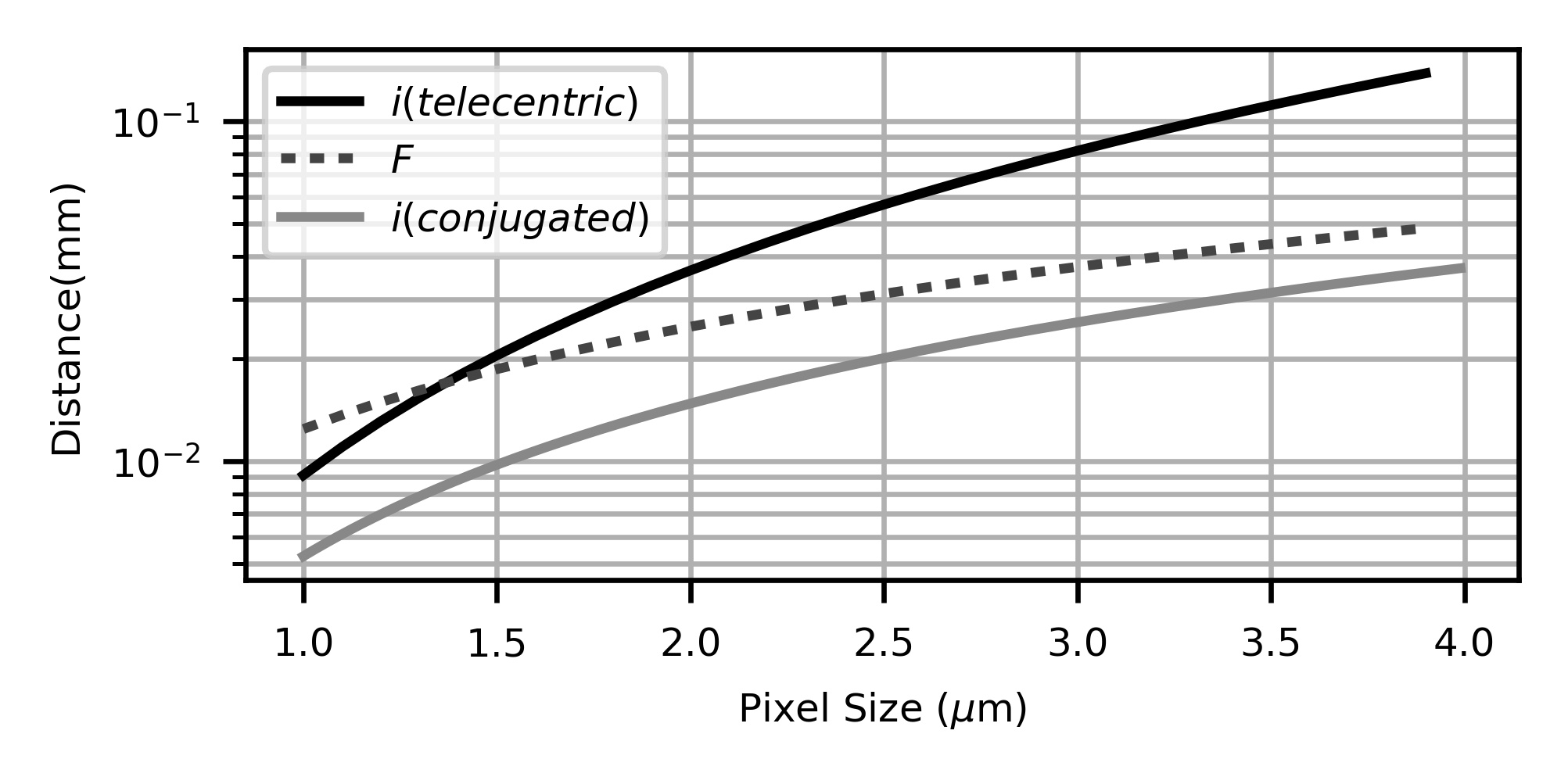}}
\caption{Evolution of the Imaging lens focal length (F) and Lenslet to imaging lens distance (i) versus the pixel size of the sensor for a 7-types hexagonal lenslet array when the eye lens and Imaging lens are conjugated and for the telecentric configuration.}
\label{fig:IandFvspixel_v1}
\end{figure}

We present a telecentric configuration in this letter. Equation\,\ref{eq:ulambda vmin} shows that for a given pixel size of the sensor and target resolution the diameter of the lenslet is fixed and evolve linearly with the pixel size of the sensor. Using equation\,\ref{eq:minimizer} and $v_{min}=\nicefrac{i}{b}$, we can calculate the imaging lens image distance from the lenslets array $i$ and from equation\,\ref{eq:fnumber matching} we calculate the focal length of the imaging lens. We plot the evolution of $i$ and $F$ against the pixel size of the sensor in figure\,\ref{fig:IandFvspixel_v1}, where we can notice that for pixel size above a certain value the effective focal length of the imaging lens get smaller than $i$. For example for a 7-types hexagonal array, a sensor with a pixel size larger than $1.5\,\mu$m will require an imaging lens with a back focal length longer than its effective focal length, otherwise know as reverse telephoto or retrofocus lens~\cite{vella_extreme_2015}.


An alternative design to the telecentric configuration presented in this letter is to conjugate the eye lens and imaging lens. A unit magnification relay telescope can be use for this purpose, effectively making both imaging lens and eye lens acting as one. In this alternative configuration, equations\,\ref{eq:ulambda vmin} and \ref{eq:minimizer} become respectively:

\begin{eqnarray}
u_{\lambda}=s_{\lambda}\left(\frac{\Delta}{d}-v_{min}\right)\label{eq:ulambda_old}\\
b=-\frac{de_{0}v_{min}}{\Delta}\frac{1}{\left(\frac{\Delta}{d}-v_{min}\right)}\label{eq:ulambda_old}
\end{eqnarray}
We plot in figure\,\ref{fig:IandFvspixel_v1}, the corresponding evolution of $i$ and $F$. Here we can see that for all pixel size the effective focal length remain larger than $i$. This configuration is suboptimal in terms of matching the f-number across the whole imaging depth, but it might significant simplify the design of the imaging lens. In the plenoptic literature the increase of the effective f-number for plane located at various distance within the imaging volume has not been considered or discussed.

\begin{figure}[tbph]
\centering
\fbox{\includegraphics[width=\linewidth]{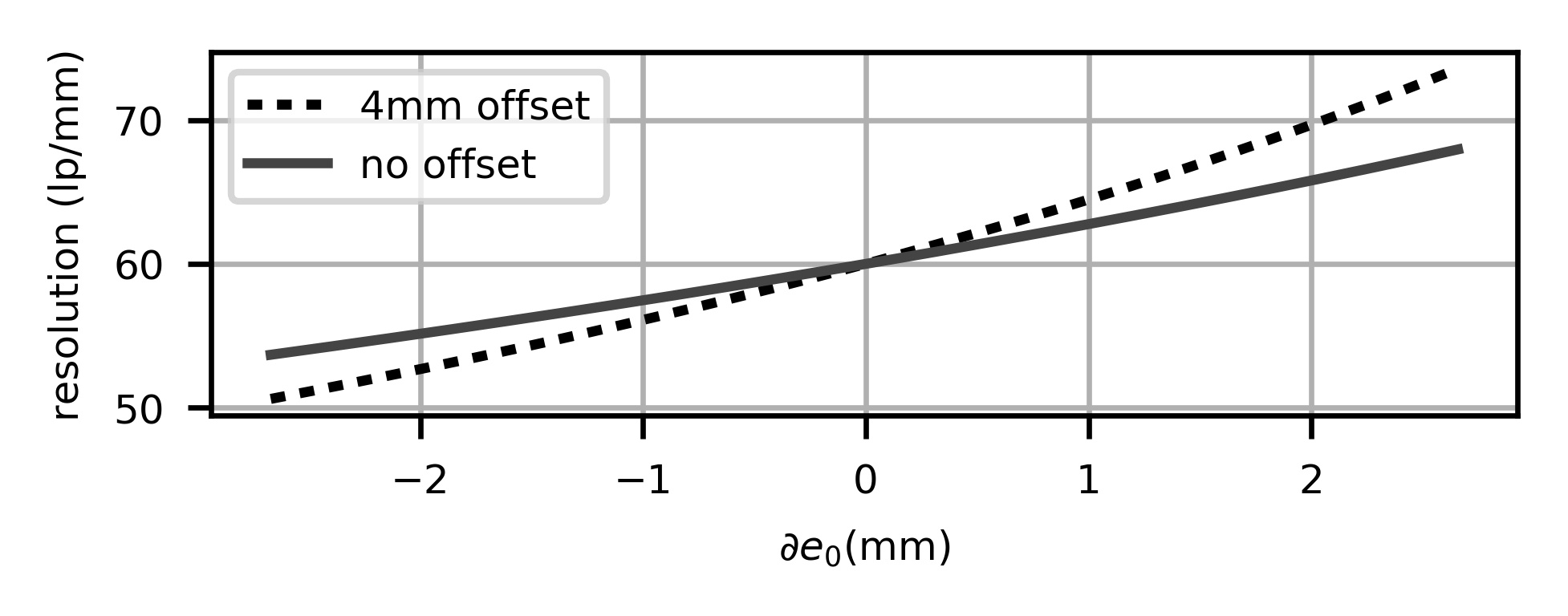}}
\caption{Effect of eyeball diameter variability across the population on the resolution of the proposed plenoptic imaging system. Resolution of the system versus the change of axial length ($\partial e_0$) for the retinal plane and a plane offset $4$\,mm within the vitreous.}
\label{fig:vitreous}
\end{figure}

The effective focal length and dimension of vitreous humour in human eyes vary among person. A recent study~\cite{kim_ultrasonographic_2017} shows a maximum variation of $5$mm across the population. We estimated the effect of this inter-subject variability on the image resolution of the proposed design by taking the derivative of equation\,\ref{eq:ulambda}. Figure\,\ref{fig:vitreous} shows the resolution change for the retinal plane and for a plane offset by $4$\,mm from the retina which corresponds to the depth extent of the imaging volume as seen in figure\,\ref{fig:7 lenslet f3 nodiff}.

\section{Conclusions}
In this letter we proposed a design approach for retinal imaging with a constant lateral resolution across an extended depth imaging range which is significantly larger than what is currently achievable with a stereo microscope without manual focusing. We established the first order properties and requirements for a plenoptic ophthalmic imaging system. The presented methodology paves the way for the implementation of such clinically-relevant systems.

\section*{Acknowledgements}
We acknowledge the support of Fight for Sight [1728/29] and the Academy of Medical Sciences [SBF001/1002]. Further, we thank Peter West (UCL Institute of Ophthalmology) for providing necessary equipment.

\bibliographystyle{plain}
\bibliography{LFO_letter}

\end{document}